\documentclass[12pt]{article}
  
\usepackage[a4paper, total={6.25in, 9.5in}]{geometry}
\usepackage[super,square]{natbib}
\usepackage{doc}
\usepackage{subcaption}
\usepackage{url}
\usepackage{sectsty}
\sectionfont{\fontsize{16}{15}\selectfont}
\subsectionfont{\fontsize{15}{15}\selectfont}
\allsectionsfont{\mdseries}

\usepackage{graphicx}
\usepackage{epstopdf}
\title{\Large An Investigation into the Kinetics and Mechanism of Phase Transitions in Optical Phase Change Ternary Alloy $Ge_2Sb_2Te_5$ }
\author{Aditya Muralidharan}
\date{}

\begin{document}

\maketitle

\abstract{
Ternary alloys of \textit{Ge-Sb-Te (GST)} have been extensively studied due to their unique ability display a reversible change in their phase upon stimulation by optical pulses i.e amorphous \textit{(a-GST)} to crystalline \textit{(c-GST)} and vice-versa. The two phases exhibit remarkably different electrical and optical properties like conductivity, reflectivity, refractive index and optical loss, this coupled with their high phase switching speeds, low power phase switching, large switching cycles, large measurable optical and electrical contrast, and phase stability makes GST alloys stand out from other phase change materials \textit{(PCM)}. GST alloys have already found extensive use in optical disks and electronic memories due to their non-volatility and zero static power consumption, but the precise mechanism of the phase change is not clearly understood. The phase change mechanism has usually been attributed to the optical pulse, usually a high power short pulse laser, heating up the \textit{c-GST} alloy to above its melting temperature (\textit{$T_m$}) after which, if it is cooled rapidly enough to below the glass transition temperature  (\textit{$T_g$}), the atoms are fixed in place due to the drastic reduction in their mobility, resulting in a phase which exhibits structure similar to a frozen liquid and lacks long range order i.e amorphous. If alternatively the \textit{a-GST} is heated above \textit{$T_g$} with a intermediate power laser pulse for a significant amount of time to induce nucleation, then this favours the shift back to the energetically favourable crystalline phase. With the growing interest in next generation data storage technology for photonic and neuromorphic computing, the research into understanding and improving the properties of GST alloys has also been rekindled. In this term paper we hope to investigate and gain a deeper understanding and appreciation of the kinetics and underlying atomistic mechanism of this phase transition from \textit{c-GST} to \textit{a-GST} and vice-versa which is only now becoming clearer. 

\pagebreak

\pagebreak

\pagebreak

\pagebreak

%
%
\section*{\Large \normalfont Introduction}
\label{introduction}
The properties of materials, especially those used in the electronics industry like silicon or \textit{$SiO_2$} are thought of as static, usually fixed at the time of device fabrication.\cite{gong}\cite{shaolin} This behaviour is usually desirable to ensure the stability and reliability of the device under variable operating conditions. But with the emergence of new paradigms like photonic and neuromorphic computing, it has become imperative to look for more dynamic material systems which go beyond this passive behaviour and can respond more robustly to optical stimulation, to achieve the ultimate goal of dynamically tunable and reconfigurable optical and electronic systems.\cite{katrin}\cite{miller}\cite{chou}

Phase change materials \textit{(PCM)} represent a class of solids which undergo a crystallographic phase transformation due to a change in their temperature compared to the ambient, usually caused by electrical or optical stimulation. \cite{lee}\cite{loke} This transition in their phase is also usually accompanied by a stark contrast in the electrical and optical properties of the two phases which is exploited for data storage applications.\cite{yifei}\cite{matthew} They can be see as stimuli responsive materials that modulate their electrical and optical behaviour in response to the external stimuli, usually a electrical potential or a laser pulse.\cite{lencer}\cite{pengfei} One of the two phases is characterized as having low reflectivity and  refractive index, low dielectric constant and high resistivity i.e. they behave like semiconductors, while the other phase shows a high reflectivity and refractive index, high dielectric constant and low resistivity i.e. they show metallic behaviour.\cite{kalb}\cite{siegert}

Ternary alloys of \textit{Ge-Sb-Te (GST)} as shown in Figure 1(b) have long shown promise for such phase change behaviour due to the existence of a metastable crystalline phase in the rock-salt structure \textit{(FCC)} which amorphizes upon stimulation by a high-power short-width laser pulse. \cite{tsafack}\cite{cho}\cite{khulbe} This transition though is reversible, as upon stimulation of the \textit{a-GST} by a intermediate-power long-width laser pulse the \textit{a-GST} recrystallizes into its metastable crystalline form.\cite{chenbin}\cite{arrigo}\cite{coombs}\cite{chu} The crystalline to amorphous transition of GST requires the \textit{c-GST} to be heated above its melting temperature \textit{$T_m$} and rapidly quenched in order to fix the atoms in place before they have a chance to migrate back to their equilibrium position, thus the resulting \textit{a-GST} structure is analogous to a "frozen liquid" with its atoms frozen in place. \cite{kalb}\cite{welnic}}\cite{schumacher}  The reverse transition on the other hand only heats the \textit{a-GST} to above its glass transition temperature \textit{$T_g$} which is maintained long enough to induce nucleation and crystallization, thus allowing it to transition back to \textit{c-GST}.\cite{mao}\cite{gonzalez} We can also infer from the phase diagram that different compositions of the ternary alloy are possible as we travel across the line from $GeTe$ to $Sb_2Te_3$ such as $GeSb_2Te_4$ and $Ge_2Sb_2Te_5$.\cite{kalb}\cite{sajjad} We also observe that the transition time between phases ($t$) and the stability of the amorphous phase along with $T_m$ and $T_g$ decrease as we move in the same direction.\cite{sajjad}\cite{delaney}\cite{raoux} $Ge_2Sb_2Te_5$ $(GST-225)$ is of particular interest in this regard, as of all the alloys on the ternary phase diagram shown in Figure 1(b), it shows the optimum position for both fast phase transition \textit{(t $<$ 20ns)} and phase stability, while having a workable $T_m \sim 650^oC$ and $T_g \sim 100^oC-150^oC$. \cite{sajjad}\cite{luong} The amorphous phase of $Ge_2Sb_2Te_5$ also has a degradation lifetime $t_d \sim 10$ years making it suitable for use in next generation non-volatile memory. \cite{shportko}\cite{irene}\cite{wang}\cite{kolobov}

\begin{figure}[htb]
\centering
\begin{subfigure}[b]{0.45\textwidth}
\centering
\includegraphics[width=2in]{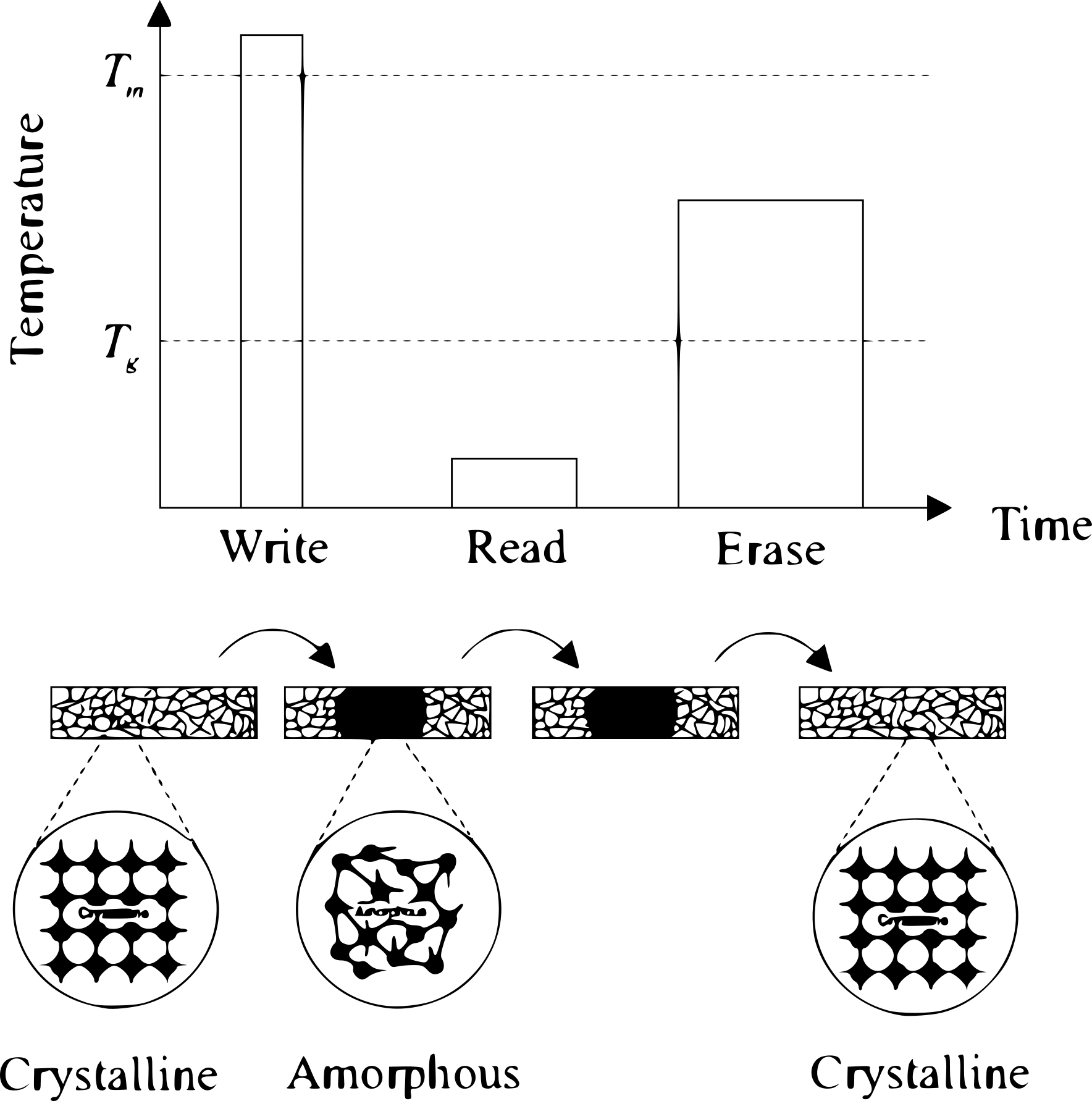}
\caption{{}}
\end{subfigure}
\begin{subfigure}[b]{0.45\textwidth}
\centering
\includegraphics[width=2in]{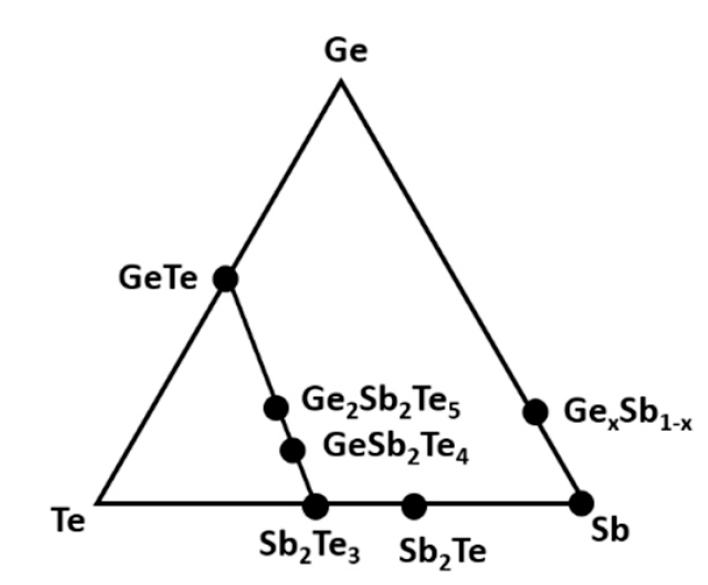}
\caption{{}}
\end{subfigure}

\caption{\textbf{\textit{(a)}} The process of amorphization and recrystallization in a phase change material using a optical pulse. The material is amorphized by the action of a high-power short-duration laser pulse that rapidly heats the material to above $T_m$ and quenches it below $T_g$. The material is then recrystallized using a intermediate-power longer-duration laser pulse that keeps it at a temperature above $T_g$. A low power laser pulse can be used to measure the reflectivity of the two phases without altering them thus allowing us to read the stored information.\cite{kalb} \textbf{\textit{(b)}} Ternary phase diagram for the Ge-Sb-Te system.\cite{wang} }
\end{figure}

In this term paper we investigate the kinetics and mechanism of such phase transitions in $GST-225$. We will first look at the stability of the two phases of $GST-225$ involved in the transition, namely the crystalline rock-salt phase $(c-GST)$ and the amorphous phase $(a-GST)$, we will also briefly touch upon the hexagonal phase $(h-GST)$.Then we will move on to study how various factors such as laser power, pulse width, quench time \& cooling rate affect both the $(c \rightarrow a)$ and ($a \rightarrow c$) transitions and the macroscopic kinetics and mechanism of phase change namely nucleation and rim to centre growth. We will also investigate the atomistic mechanism of the phase change and look at bond distortions in the structure which when coupled with atomic migrations and carrier collisions facilitate phase change in $GST-225$ upon stimulation.  

\begin{figure}[t]
\centering
\begin{subfigure}[b]{0.45\textwidth}
\centering
\includegraphics[width=1.5in]{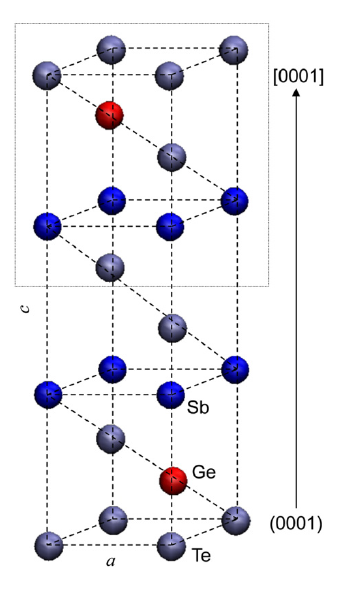}
\caption{{}}
\end{subfigure}
\begin{subfigure}[b]{0.45\textwidth}
\includegraphics[width=1.60in]{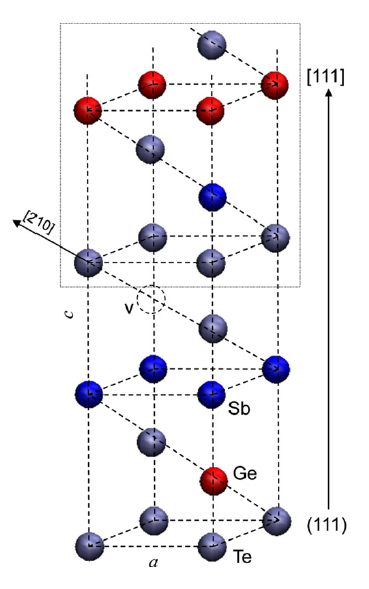}
\centering
\caption{{}}
\end{subfigure}
\caption{\textbf{\textit{(a)}} hexagonal-GST \textbf{\textit{(b)}} rock-salt cubic-GST.\cite{tsafack}}
\end{figure}

\section*{\normalfont $Ge_2Sb_2Te_5$}
$Ge_2Sb_2Te_5$ $ (GST-225)$ is a ternary alloy in the Ge-Sb-Te system which contains Germanium, Antimony and Tellurium in a stoichiometric ratio of $2:2:5$, being roughly in the middle of the $GeTe-Sb_2Te_3$ line in the ternary phase diagram. \cite{sajjad} $GST-225$ is an excellent PCM that is of immense interest due to its splendid set of properties, namely high thermal stability,fast crystallization speed non-volatility and reliability. \cite{sajjad}\cite{raoux} 

\begin{figure}[t]
\centering
\begin{subfigure}[b]{0.45\textwidth}
\centering
\includegraphics[width=1.6in]{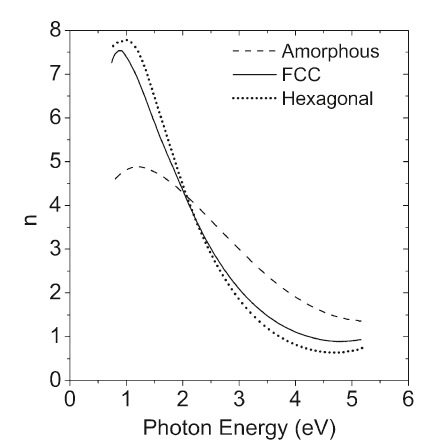}
\caption{{}}
\end{subfigure}
\begin{subfigure}[b]{0.45\textwidth}
\includegraphics[width=1.60in]{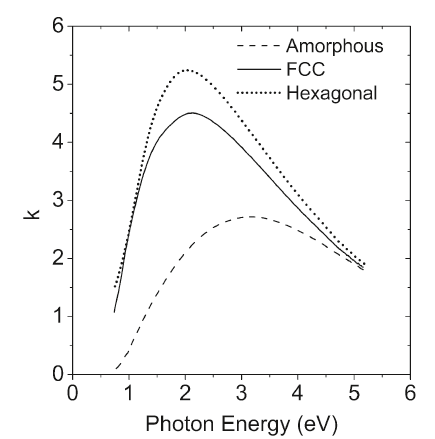}
\centering
\caption{{}}
\end{subfigure}
\begin{subfigure}[b]{0.45\textwidth}
\includegraphics[width=1.60in]{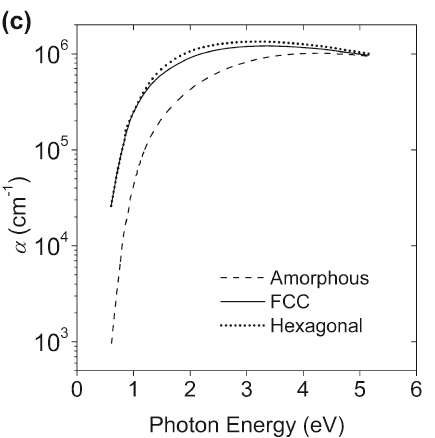}
\centering
\caption{{}}
\end{subfigure}
\begin{subfigure}[b]{0.45\textwidth}
\includegraphics[width=2.4in]{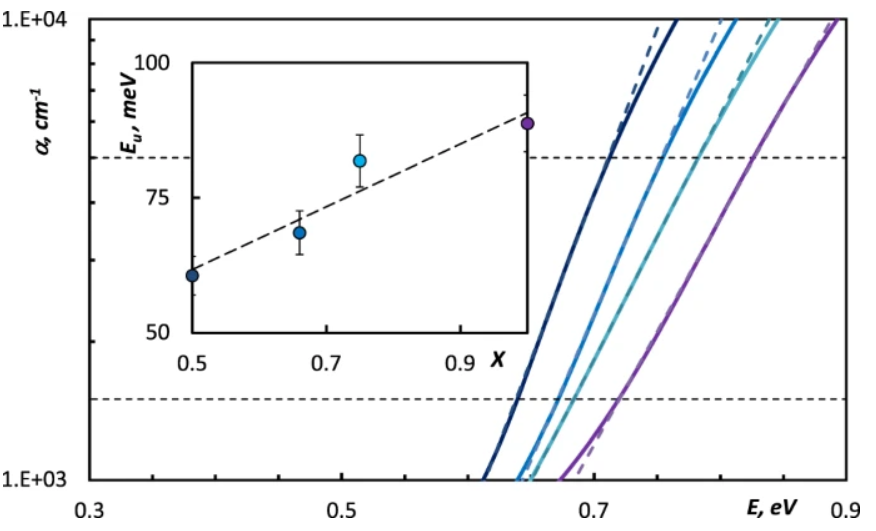}
\centering
\caption{{}}
\end{subfigure}
\caption{\textbf{\textit{(a)}} real part of the refractive index $(n)$ \textbf{\textit{(b)}} imaginary part of the refractive index $(k)$ \textbf{\textit{(c)}} absorption coefficient for $GST-225$\cite{lee}\textbf{\textit{(ds)}} Urbach-Martiensenn tail in the absorption coefficient of $GST-225$\cite{shportko}}
\end{figure}

\subsection*{\normalfont Crystalline $Ge_2Sb_2Te_5$}
Such a impressive set of properties is partly due to the structural arrangement of atoms in $GST-225$ which crystallizes in two phases, the metastable cubic rock-salt phase $(c-GST)$ and more energetically favourable hexagonal phase $(h-GST)$ as seen in Figure 2. \cite{tsafack}\cite{sajjad} As deposited films of $a-GST$ using magnetron sputtering are observed to transition to $(c-GST)$ at $\sim 150^oC$ and again to $(h-GST)$ at $\sim 260^oC$ indicating that the actual order of phase transitions is $(a-GST) \rightarrow (c-GST) \rightarrow (h-GST)$. \cite{wei} The reason that the metastable $c-GST$ phase is preferred for most applications is twofold. First, the activation energy of the $(a-GST) \rightarrow (c-GST)$ is of the order of $\sim 2-3 eV$ and an effective activation energy of $\sim 0.29-0.33 eV$, which is considerably lower than the activation energy and effective activation energy for the $(a-GST) \rightarrow (h-GST)$ transition.\cite{gonzalez}\cite{muneer} Second, the $c-GST$ is a metastable phase between $a-GST$ and $h-GST$ thus is easier to reverse back to the amorphous phase from $c-GST$ than the more energetically favourable $h-GST$. \cite{muneer}\cite{orava} It can also be seen from Figure 3 that the change in the optical properties such as the real and imaginary part of the refractive index $(n)$ and $(k)$ as well as the absorption coefficient $(\alpha)$ is more striking between the $a-GST$ and $c-GST$, and less between $c-GST$ and $h-GST$. \cite{lee}\cite{bragaglia}\cite{bin} This optical contrast is primarily due to atomic migration of the $Ge/Sb$ atoms during phase transition and the associated change in the band structure which creates differences in the band gaps $(E_g)$ of the $c-GST (E_g \sim 0.56eV$ with a indirect component of $E_g \sim 0.1eV)$, $h-GST (E_g \sim 0.2eV)$ and the $(a-GST E_g \sim 0.68 eV)$. \cite{kolobov}\cite{kellner} Notice, how $a-GST$ also shows a significant Urbach-Martiensenn tail in its absorption coefficient compared to $c-GST$ and $a-GST$, this is primarily due to the lack of long range order, presence of carriers near the band edge and dangling bonds caused by localized vacancies as shown in Figure 3(d). \cite{shportko}\cite{li}\cite{jiang}

\begin{figure}[t]
\centering
\begin{subfigure}[b]{0.45\textwidth}
\centering
\includegraphics[width=2.6in]{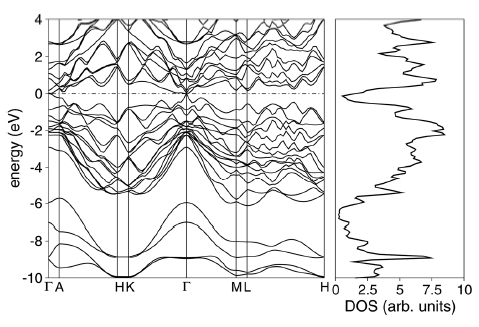}
\caption{{}}
\end{subfigure}
\begin{subfigure}[b]{0.45\textwidth}
\includegraphics[width=2.6in]{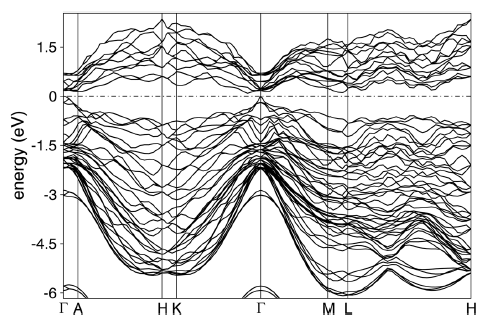}
\centering
\caption{{}}
\end{subfigure}
\begin{subfigure}[b]{0.45\textwidth}
\includegraphics[width=2in]{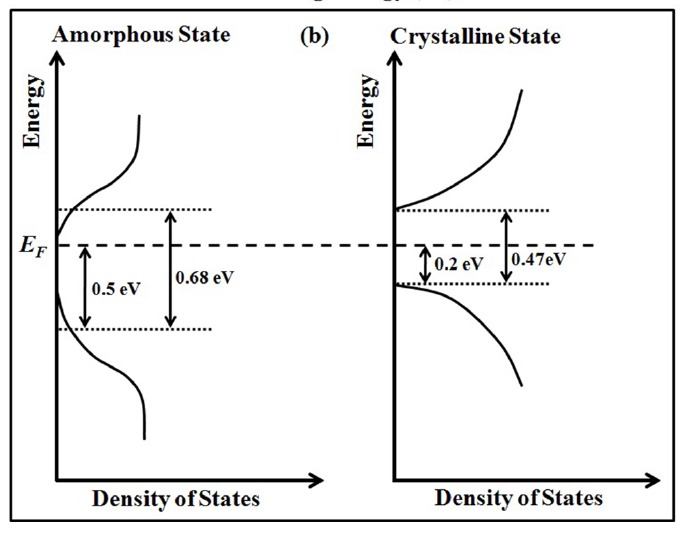}
\centering
\caption{{}}
\end{subfigure}
\caption{\textbf{\textit{(a)}} Band-gap of Hexagonal $GST$ \textbf{\textit{(b)}} Band-gap of Cubic $GST$ \textbf{\textit{(c)}} absorption coefficient for $GST-225$ \cite{tsafack} \textbf{\textit{(c)}} Simplified band gap diagrams for $a-GST$ and $c-GST$.\cite{singh}}
\end{figure}

\subsubsection*{\normalfont Hexagonal $Ge_2Sb_2Te_5$}
Hexagonal $GST$ $(h-GST)$ is the most energetically stable phase of $GST-225$, with its atoms arranged in a hexagonal close packed structure with nine atoms arranged in a stacked layered structure in each unit cell with one of the possible orders being $(Te - Ge - Te - Sb - Te - Te - Sb - Te - Ge)$ as shown in Figure 2(a) .\cite{tsafack}\cite{singh} The lattice parameters for the hexagonal unit cell have been experimentally determined to be $a=4.22$ \AA{}  and $c=17.18$ \AA{}, leading to a larger $Te-Te$ bond length of $\sim 3.7$\AA {} compared to the $Ge-Te$ and $Sb-Te$ bond length of $\sim 3$\AA{.\cite{tsafack}\cite{pan}}   
Additionally Density Functional Theory $(DFT)$ has been used in studies to calculate the band structures of both $h-GST$ and $c-GST$ in order to explain their electrical and optical properties as shown in Figure 4. \cite{kellner}\cite{zhangw} These studies calculate the band gap of $h-GST$ to be $\sim 0.2 eV$, thus predicting its behaviour as a semi-metal, which has been experimentally verified. \cite{kellner}\cite{singh} They have also proven instrumental in explaining the contrast in the optical properties between the phases, such as the refractive index contrast $(\Delta n)$ and the contrast in the spectral transmittance between the different phases $((a-GST)$ $T \sim 92$\%$)$, $((c-GST)$ $T \sim 46$\%$)$ and $((h-GST)$ $T \sim 2$\%$)$ at $\lambda \sim 2740 nm$ and the Near-IR region in general.\cite{singh} These can be explained by the change in the shape of the valence band as well as the density of states after the phase transition.\cite{kellner}\cite{yang}\cite{song} $DFT$ studies have also helped us explain the low thermal conductivity in $h-GST$ $(\kappa=1.76 W/m.K)$ as well as the large electronic component to thermal conductivity $(\kappa_e \sim 0.46W/m.K)$ when compared to $a-GST$  by calculating the electronic band structures,  phonon dispersion curves and density of states as shown in Figure 5. \cite{pan}\cite{mukhopadhyay}\cite{zhitang}\cite{xu}
The $h-GST$ phase is also known to have lower vacancy densities, $Ge-Te$ bond distortion and a higher resistance towards disorder when compared to the $c-GST$ phase which partly contributes to its stability, but also makes the transformation to $a-GST$ more difficult. \cite{cho}\cite{bragaglia}\cite{bin}\cite{kellner} This is due to the fact that the vacancies in the $h-GST$ have to arrange themselves into highly ordered layers $(super-lattice$ $layers)$ when sufficient energy is provided to the stable $hcp$ structure to show a metal to insulator transformation $(IMT)$.\cite{bragaglia}\cite{yang} This is due to a weak Van der Walls force between the low-coordinated $Te$ atomic layers between the highly ordered  vacancy layers which results in the formation of a Van der Walls gap, generally thought of as responsible for the metallic conduction and behaviour of the $h-GST$ as a degenerate p-type semiconductor.\cite{hafermann}
It has also been hypothesized that disorder induced $(Anderson)$ localizations play a pivotal role in the IMT process, making the the electrical and optical properties of the phase strongly dependent on the degree of disorder within the material.\cite{zhangw}\cite{hafermann}
Thus the central role that vacancies and bond distortions play in theses $Ge/Sb$ atomic migrations that cause the $(h-GST \rightarrow a-GST)$ and $(c-GST \rightarrow a-GST)$ phase changes is studied in following sections.

\begin{figure}[t]
\centering
\begin{subfigure}[b]{0.3\textwidth}
\centering
\includegraphics[width=1.77in]{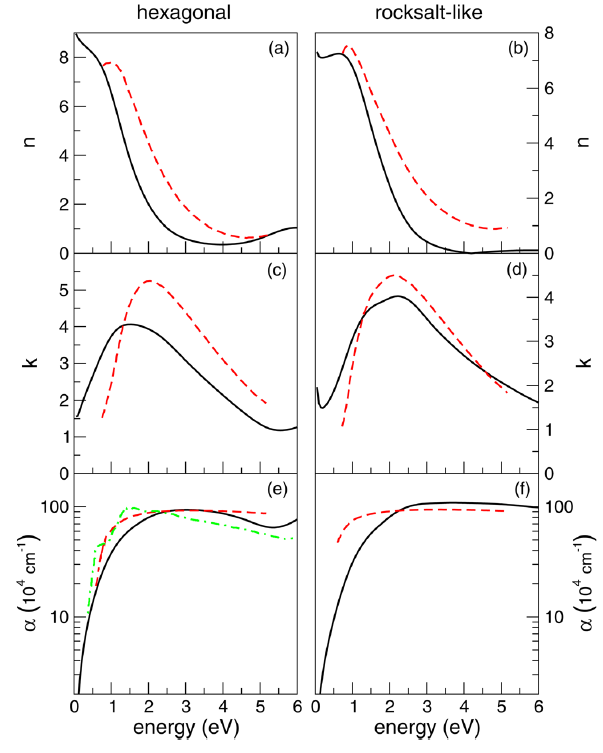}
\caption{{}}
\end{subfigure}
\begin{subfigure}[b]{0.3\textwidth}
\includegraphics[width=1.6in]{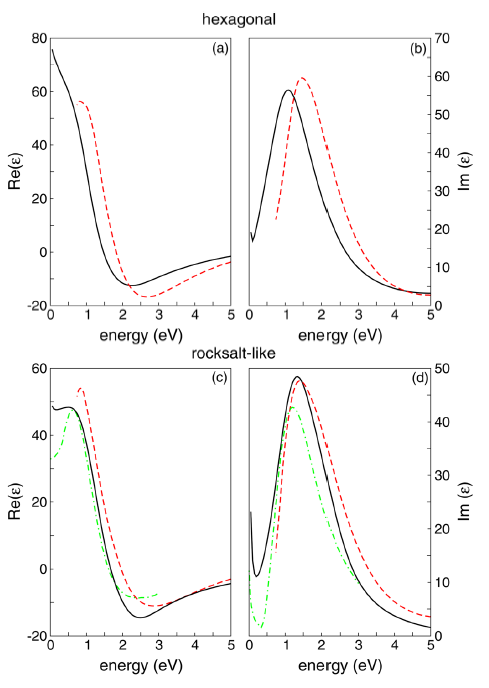}
\centering
\caption{{}}
\end{subfigure}
\begin{subfigure}[b]{0.3\textwidth}
\includegraphics[width=1.6in]{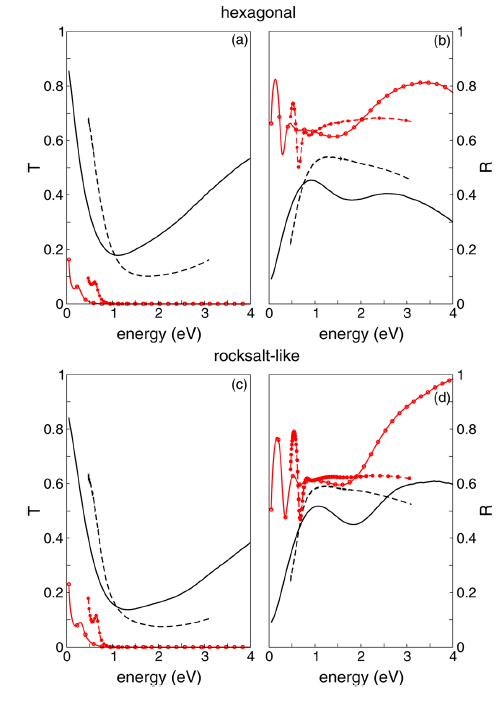}
\centering
\caption{{}}
\end{subfigure}
\begin{subfigure}[b]{0.30\textwidth}
\includegraphics[width=1.3in]{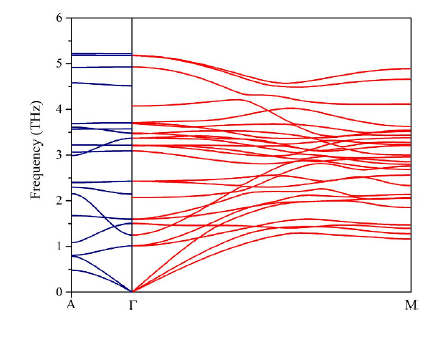}
\centering
\caption{{}}
\end{subfigure}
\begin{subfigure}[b]{0.30\textwidth}
\includegraphics[width=1.3in]{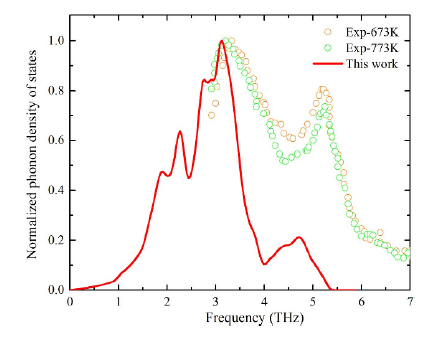}
\centering
\caption{{}}
\end{subfigure}
\begin{subfigure}[b]{0.30\textwidth}
\includegraphics[width=1.3in]{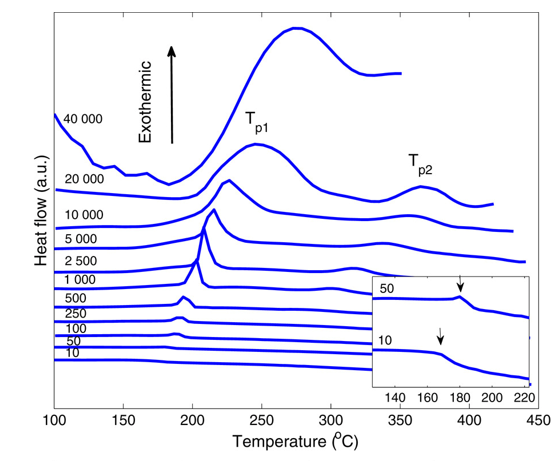}
\centering
\caption{{}}
\end{subfigure}

\caption{Comparison of DFT and Experimental Data $(dotted$ $lines)$ \textbf{\textit{(a)}} $n, k$ and $\alpha$ \textbf{\textit{(b)}} Real and imaginary parts of dielectric $\epsilon_{re}  $ and $\epsilon_{im}$ \textbf{\textit{(c)}} Transmittance $T$ and Reflectance $R$\cite{tsafack}
\textbf{\textit{(d)}} phonon dispersion relation \textbf{\textit{(e)}} phonon density of states\cite{mukhopadhyay}
\textbf{\textit{(f)}} Ultra-fast DSC measurements of $GST$ nano-particles.\cite{chenbin}
}
\end{figure}

\begin{figure}[htp]
\centering
\begin{subfigure}[b]{1\textwidth}
\centering
\includegraphics[width=5in]{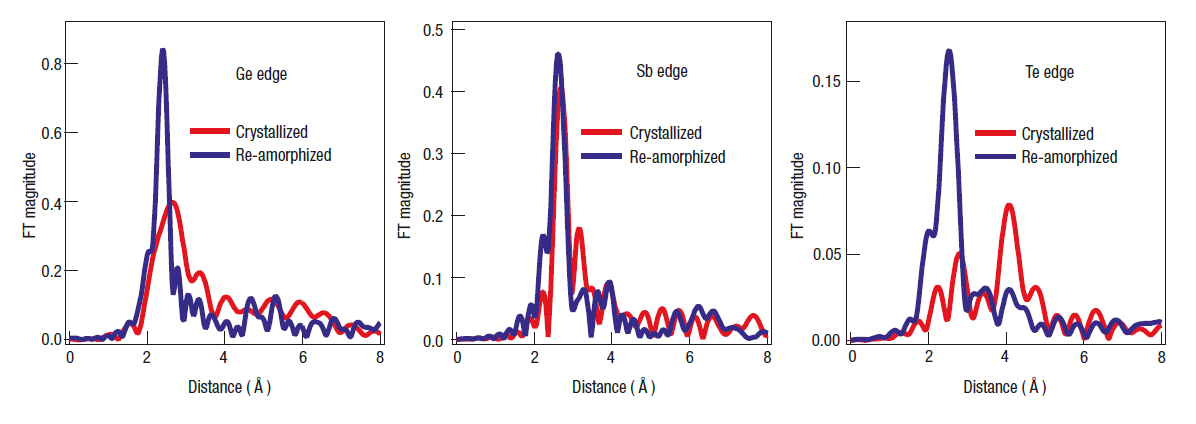}
\caption{{}}
\end{subfigure}
\begin{subfigure}[t]{0.3\textwidth}
\centering
\includegraphics[width=1in]{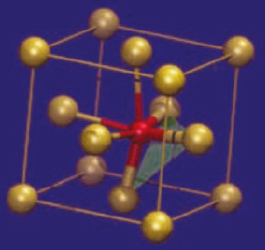}
\includegraphics[width=1in]{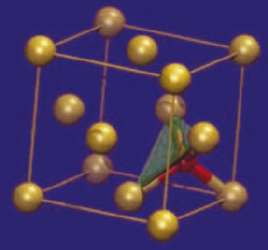}
\caption{{}}
\end{subfigure}
\begin{subfigure}[t]{0.3\textwidth}
\centering
\includegraphics[width=1.4in]{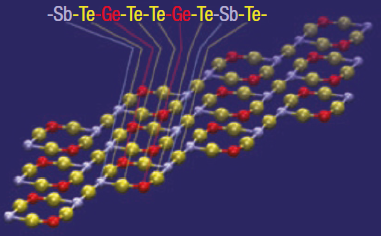}
\caption{{}}
\end{subfigure}
\begin{subfigure}[t]{0.3\textwidth}
\centering
\includegraphics[width=1.4in]{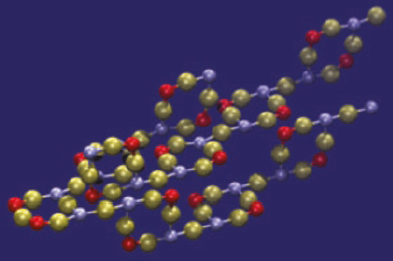}
\caption{{}}
\end{subfigure}
\begin{subfigure}[t]{1\textwidth}
\centering
\includegraphics[width=1.3in]{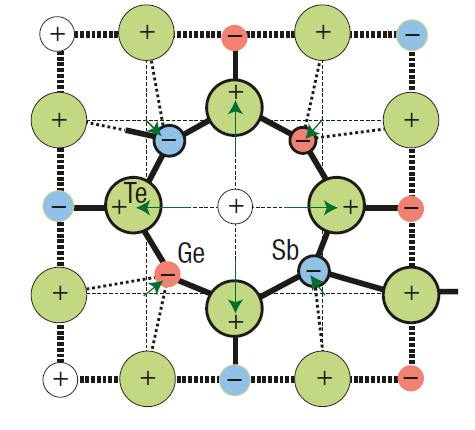}
\caption{{}}
\end{subfigure}

\caption{\textbf{\textit{(a)}} Fourier Transform EAXFS measurement for $c-GST$. The bond length $L_b$ is measured by studying the position of the peaks. It can be observed that the peaks for $Ge/Sb$ both show a double maxima at the distances corresponding to their bond length., indicating that the $Ge-Te$ and $Ge-Sb$ have a longer $(\sim 3.20$\AA$)$ and shorter $(\sim 2.84$\AA$)$ bond. The other bond lengths, namely $(Te-Te)$ $FCC$ $(\sim 4.26$\AA$)$ bond and $(Te-Te)$ lattice parameter bond $(\sim 6.02$\AA$)$ are extracted similarly.  \textbf{\textit{(b)}} Unit cell of the distorted Rock-salt cubic-$GST$ structure and the amorphous-$GST$. 
\textbf{\textit{(c)}} layer stacking in $c-GST$ \textbf{\textit{(d)}} $a-GST$
\textbf{\textit{(e)}} Frontal view of a unit cell with a vacancy in the octahedral site.\cite{kolobov}}
\end{figure}

\subsubsection*{\normalfont Face Centered Cubic $Ge_2Sb_2Te_5$}

$GST-225$ also crystallizes in a rock-salt type structure with the $Te$ atoms occupying the Face Centred Cubic $(FCC)$ lattice while the $Ge$ and $Sb$ atoms and vacancies occupy the octahedral voids in the body centre as shown in Figure 6(e). \cite{sajjad}\cite{kolobov} It can also be seen that the equilibrium position for the atom at the body centre of the $c-GST$ unit cell is deviated towards one of the quadrants, this is confirmed by both XRD and EAXFS measurements of the lattice parameter $(a \sim 6.02$\AA$)$ and the bond lengths of the $Ge-Te$ and $Sb-Te$ bonds as seen in Figure 6(a) \& 8(a). \cite{kolobov} It should also be mentioned that even though the isotropic displacement of the  $Ge$ and $ Sb$ atoms  $<u^2>$ are large $\sim 0.04$ \AA$^2$, the mean square relative displacement obtained by EAXFS of the $Ge-Te$ and the $Sb-Te$ is small $\sim 0.02$\AA$^2$.\cite{kolobov} This definitively implies that the structure is not an ideal but a distorted rock-salt arrangement where the direction of the distortion is not perfectly random but is heavily influenced by the surrounding $Te$ atoms. \cite{kolobov}\cite{yang} The reason for this distortion is thought to be due to the differences in the covalent radii of the $Ge/Sb$ atoms, mainly $Ge$, which is deviated from the perfect $FCC$ centre resulting in a system with shorter $(L_b\sim 2.84$\AA$)$ and longer $(L_b\sim 3.20$\AA$)$ $Ge-Te$ bonds, and $Te-Te$ bonds$(L_b \sim 4.26$\AA$)$ in the same unit cell, thus forming an overall buckled system. \cite{kolobov}\cite{yang}\cite{xu} The shorter bonds are more rigid than the longer bonds thus bringing the $Ge$ atom closer to one of the corners, and three of the $FCC$ $Te$ atoms, effectively resulting in a Peierls distortion in the local structure as seen in Figure 6(b) .\cite{cho}\cite{kolobov}\cite{zhangw}

In an ideal $FCC$ cubic structure every atom needs to form 6 bonds due to inherent octahedral symmetry, this implies each atom must contribute $6e^-$ to bonding. The $Ge$ atoms and $Sb$ atoms having only $4e^-$ and $5e^-$ in their valence shell respectively are thus in need of excess electrons, which are provided by $Te$ atoms adjacent to octahedral vacancies.\cite{kolobov} This strongly suggests that the presence of vacancies is an intrinsically important part of the structure. Experiments in which the concentrations of $Ge/Sb$ atoms were increased showed that the excess $Ge/Sb$ did not occupy the octahedral vacancies but accumulates at the grain boundary.\cite{wang}\cite{kolobov}\cite{bragaglia}\cite{bin}\cite{zhangw} It is proposed that this is due to the need for vacancies to maintain the long range order in the crystal structure and thus maintain crystal stability.\cite{kolobov}\cite{bragaglia}\cite{bin} It is also proposed that vacancies are usually formed at the sites that would have been occupied by the $Ge$ atoms, which has been experimentally verified and reported.\cite{bragaglia}\cite{yang} Studies have reported formation of upto $10\%$ $-$ $25\%$ of the vacancies in $Ge$ occupied sites upon the thermal crystallization of as-deposited $GeTe$ films. The ideal crystal structure of $c-GST$ is shown in Figure 6(c), where only the shorter $Ge-Te$ and $Ge-Sb$ bonds are shown for simplicity.\cite{welnic}\cite{kolobov}\cite{zhangw}

The real crystal structure of $GST-225$ deviates from the ideal in two respects, namely that $Ge/Sb$ atoms may swap places resulting in randomness and that each additional unit cell may rotate $90^o$ arbitrarily.\cite{kolobov}\cite{sun} This results in the $Te$ atoms forming a $FCC$ sub-lattice and $Ge/Sb$ atoms and vacancies in the other $FCC$ sub-lattice. An important point to note is that even though the $Ge/Sb$ vacancies are distributed randomly in the long range, in the short range they form well defined rigid blocks.\cite{chenbin}\cite{kolobov} It is known that $FCC$ $(-A-B-C-A-B-C-)$  and $HCP$ $(-A-B-A-B-)$ are both close packed structures with minor differences in the way that atomic layers arrange themselves, thus the ideal structure depicted by Figure 6(c) seen as a ordered chain of 9 atomic layers $(- Sb - Te - Ge - Te - Te - Ge - Te - Sb - Te -)$ can be thought of as a prelude to the energetically favourable hexagonal arrangement.\cite{tsafack}\cite{chenbin}\cite{chen} This can be observed in the differential scanning calorimetry measurements in Figure 5(f) where an extra exothermic peak is seen between those corresponding to the crystallization of the $a-GST$ and the transformation into $h-GST$.\cite{chenbin} The band structure of $c-GST$ calculated using $DFT$ also indicates that it should behave like a semiconductor with a band gap of $\sim 0.56 eV$, consistent with experimental measurements.\cite{tsafack}\cite{kellner}

\begin{figure}[t]
\centering
\begin{subfigure}[b]{0.3\textwidth}
\centering
\includegraphics[width=1.7in]{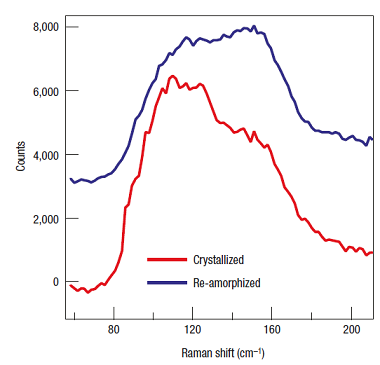}
\caption{{}}
\end{subfigure}
\begin{subfigure}[b]{0.3\textwidth}
\centering
\includegraphics[width=1.7in]{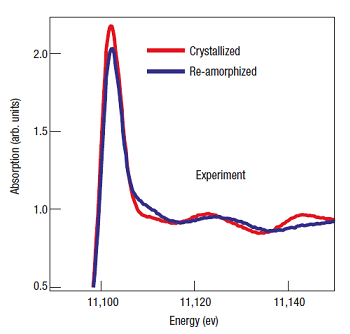}
\caption{{}}
\end{subfigure}
\begin{subfigure}[b]{0.3\textwidth}
\centering
\includegraphics[width=1.7in]{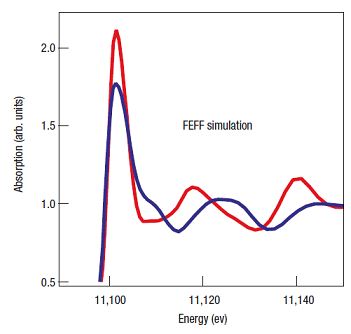}
\caption{{}}
\end{subfigure}

\caption{\textbf{\textit{(a)}} Raman spectra for the $a-GST$ and $c-GST$ phases. 
\textbf{\textit{(b)}} Measured and  \textbf{\textit{(c)}} simulated XANES spectra for $a-GST$ and $c-GST$.\cite{kolobov}}
\end{figure}

\subsection*{\normalfont Amorphous $Ge_2Sb_2Te_5$}
We now turn to the amorphous phase of $GST-225$ in which the $Ge-Te$ $(\sim 2.61$\AA$)$ and $Sb-Te$  $(\sim 2.85$\AA$)$ bonds are shorter and stronger. \cite{kolobov}\cite{arrigo} Such behaviour is unusual for such solids as the bond lengths tend to get longer upon amorphization due to the anharmonicity of the inter-atomic potential.\cite{kolobov}\cite{zhang} It has been hypothesized that rupturing of longer $Ge-Te$ bonds in $c-GST$ resulting in a system with shorter bond lengths is responsible for the effect.\cite{kolobov}\cite{zhang} This may well be the case, as the existence of long and short $Ge-Te$ and $Sb-Te$ bonds is experimentally verified in the distorted structure of the $c-GST$ .\cite{kolobov} The local order present in the $a-GST$ structure is also investigated with the help of Raman scattering where the Raman modes measured for $a-GST$ were more rigid than the crystalline state as seen in Figure 7(a), which supports the theory of bond rupture .\cite{kolobov}\cite{song} It is also observed that the volume of $a-GST$ decreases upon crystallization which is due to the weakening of the intermolecular interactions caused by the same ruptured bonds.\cite{kolobov} XANES measurements and simulations for the $a-GST$ structure also indicate that the $Ge$ atom was allowed to acquire its favoured tetrahedral arrangement, only possible with the rupture of the longer bonds as discussed in section $4.1$.\cite{kolobov}\cite{song}\cite{zhitang}

\subsection*{\normalfont Contrast in electrical \& optical properties between c-GST \& a-GST}
The primary reason for immense interest in $GST$ phase change materials as opposed to other classes of PCM's is that they display a unusually high contrast in electrical and optical properties, like real and imaginary component of the refractive index $ (\Delta n $ and $ \Delta k )$, conductivity $(\sigma)$, transmittance $(T)$, reflectance $(R)$, and absorptivity $(\alpha)$, under and after stimulation, while also maintaining that change long after the source of the stimulation has been removed.\cite{lee}\cite{kalb}\cite{irene}\cite{kolobov} These changes in electronic and optical properties are caused by a combination of structural changes to the lattice after transformation and the band gap change associated with it.\cite{bragaglia}\cite{kellner} Quantum mechanical $DFT$ calculations using fermi's golden rule have also shown that the profound change in optical properties can be explained by such structural alterations.\cite{bragaglia}\cite{kellner}\cite{song}

Vacancies in the material, especially $Ge$ vacancies in the body centre are thought to play a major role in the materials ability to show such contrast on phase transformation due to the intimate role that vacancies have in facilitating IMT as discussed in the previous sections.\cite{welnic}\cite{bin}\cite{zhangw}\cite{hafermann}
Although the exact mechanism behind the trend is not well understood, increasing the stoichiometric concentration of $Ge$ in the lattice has shown to increase the optical contrast considerably, increasing from $(\Delta n + i\Delta k =-1.48 + i 1.35)$ for $Ge_8Sb_2Te_{11}$ to $(\Delta n + i\Delta k =-1.2 + i 1.05)$ for $Ge_2Sb_2Te_5$. The electron behaviour and the density of states in the longer and shorter $Ge-Te$ bonds is also speculated to play a role in the optical contrast displayed by these materials.\cite{kolobov}\cite{muneer}\cite{yang}

\section*{\normalfont Macroscopic Kinetics of Solidification and Phase transformation}

$GST-225$ and other phase change alloys have been used in next generation rewritable optical media for the last few years and are being considered for use in high speed non-volatile memory. The primary method for inducing this phase change, which is used for writing and rewriting bits onto a memory cell, is a $SET$ and $RESET$ laser pulse of varying power and temporal width.\cite{kalb}\cite{welnic} Therefore fast recrystallization speeds, high data transfer rate, good scalability and non-volatility of the phase transformation are major factors that make it imperative to understand and improve the kinetics of the phase transformation for out-competing existing and emerging non-volatile memories.\cite{kalb}\cite{chenbin}\cite{siegel}

Phase transformation in $GST-225$ is fundamentally in response to temperature change caused by a incident laser pulse, which in case of a $(c-GST \rightarrow a-GST)$ transformation is high-power and short pulse width, causing the temperature of the $c-GST$ to increase rapidly to above $T_m$ $(\sim 650^oC)$.\cite{kalb}\cite{qi}\cite{cai} The rapid quenching of the heated $GST$ $(\sim 1^oC/ns)$ cause atoms to freeze in place as the temperature drops below $T_g$. The $(c \rightarrow a)$ transformation usually does not fully amorphize the crystalline phase leaves behind some sub-critical crystalline domains which play an integral part in the $(a-GST \rightarrow c-GST)$ transformation.\cite{konishi}\cite{huang}

Crystallization of $a-GST$ involves the use of a intermediate-power long-pulse width which increases the temperature of the system to above $T_g$ $(\sim 150^oC)$, this allows the atoms enough wiggle room to arrange themselves in a energetically favourable position, allowing the formation of nucleation sites and subsequent crystallization of the domains.\cite{kalb}\cite{zhou}\cite{jc} The nucleation step in this process is inherently slow and the time limiting step for the phase change to occur.\cite{kalb}\cite{chenbin}\cite{han} The growth step on the other hand is significantly faster thus it is favourable for better phase switching behaviour. Crystallization of melt-quench $GST$ is different from as-deposited films, which are dominated by a nucleation driven process while melt-quench favours a growth dominated process.\cite{kalb}\cite{wang} The amorphization of $c-GST$ by melt-quenching also leaves behind sub-critical crystalline domains in the amorphous matrix, making it possible for the slow nucleation step $(t\sim 1\mu s)$ to be bypassed by significantly faster growth dominated recrystallization $(t\sim 20-30ns)$.\cite{kalb}\cite{wang} The application of a preconditioning short laser pulse $(t_p \sim 100ns)$ can help as-deposited films mimic the speed of melt-quench films.\cite{wang}

\begin{figure}[t]
\centering
\begin{subfigure}[b]{0.3\textwidth}
\centering
\includegraphics[width=1.7in]{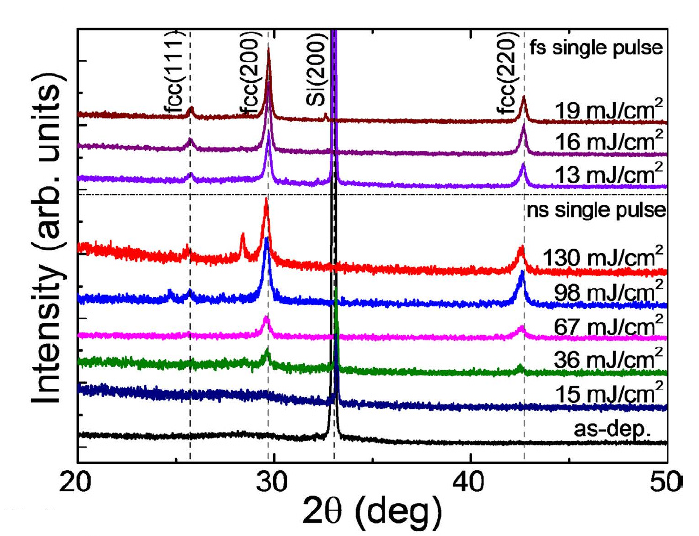}
\caption{{}}
\end{subfigure}\hspace*{\fill}
\begin{subfigure}[b]{0.3\textwidth}
\centering
\includegraphics[width=1.7in]{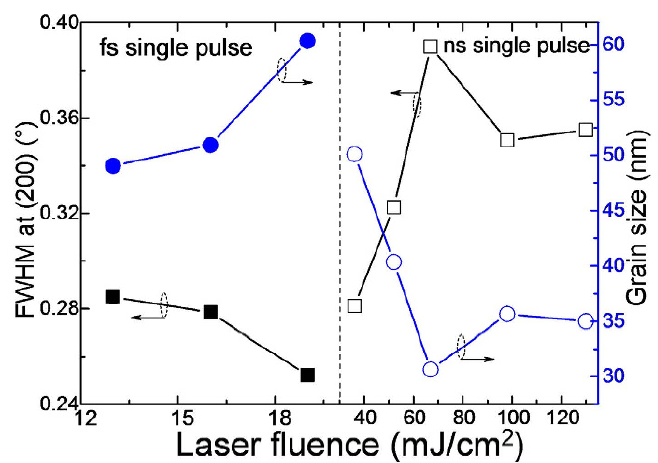}
\caption{{}}
\end{subfigure}\hspace*{\fill}
\begin{subfigure}[b]{0.3\textwidth}
\centering
\includegraphics[width=1.7in]{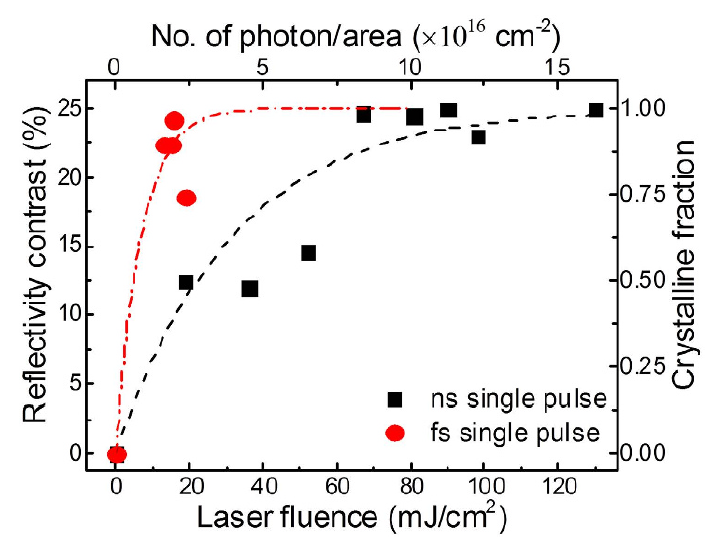}
\caption{{}}
\end{subfigure}\hspace*{\fill}

\begin{subfigure}[h]{0.3\textwidth}
\centering
\includegraphics[width=1.7in]{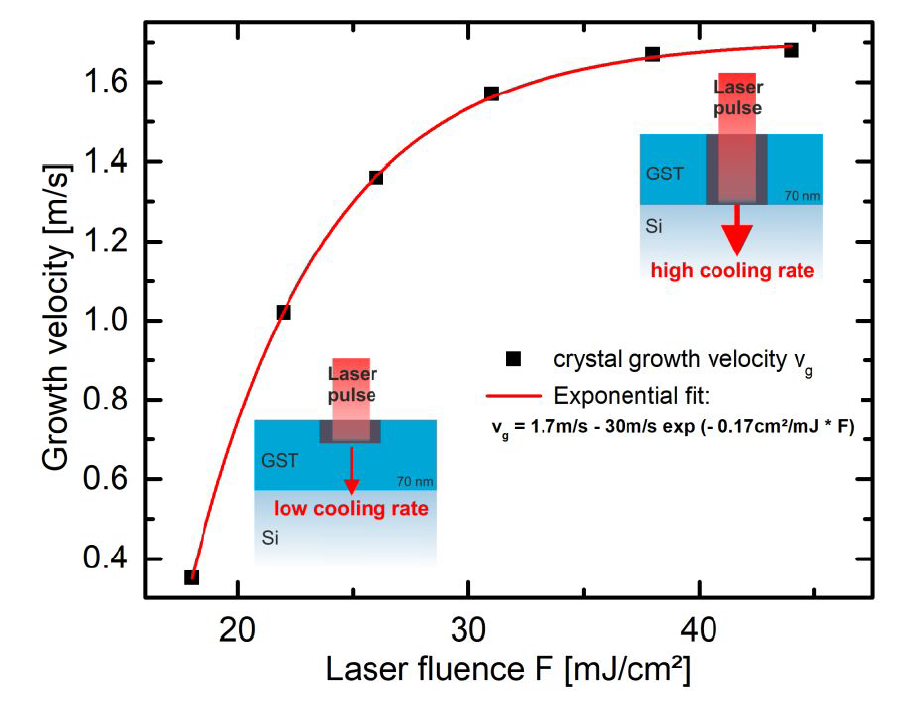}
\caption{{}}
\end{subfigure}\hspace*{\fill}
\begin{subfigure}[h]{0.3\textwidth}
\centering
\includegraphics[width=1.7in]{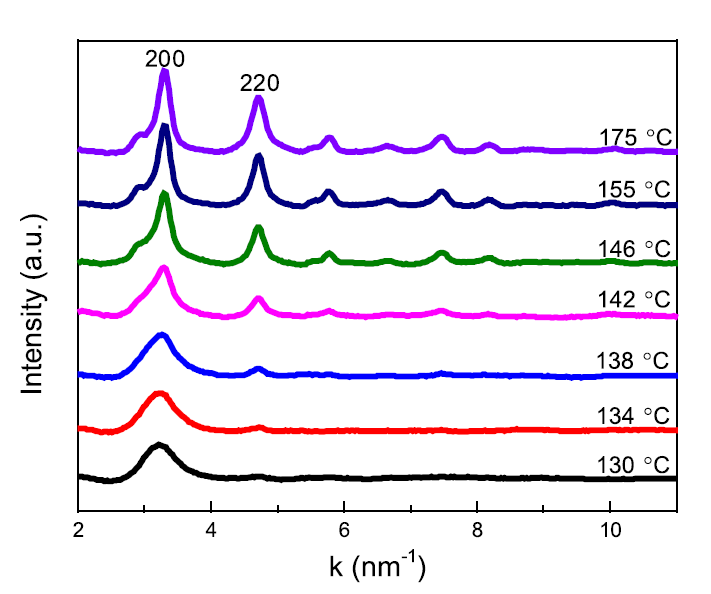}
\caption{{}}
\end{subfigure}\hspace*{\fill}
\begin{subfigure}[h]{0.3\textwidth}
\centering
\includegraphics[width=1.7in]{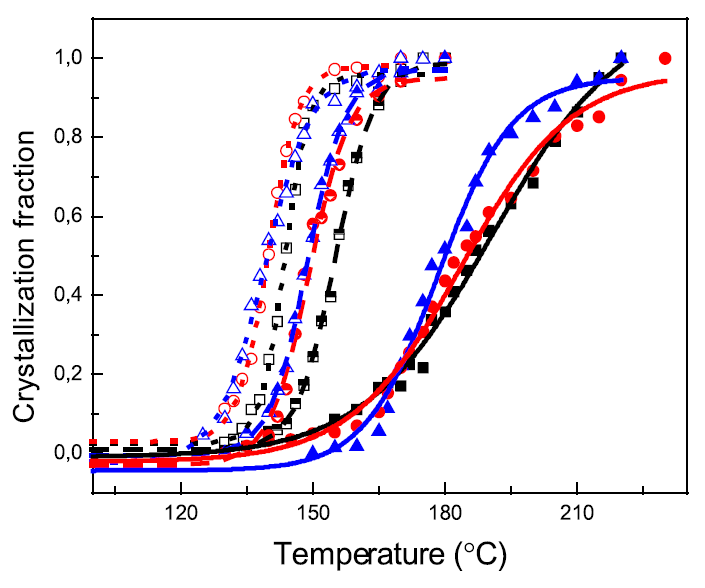}
\caption{{}}
\end{subfigure}

\caption{\textbf{\textit{(a)}} XRD spectra for $GST-225$ at different laser fluences. 
\textbf{\textit{(b)}} Grain size for $GST-225$ at different laser fluences.  \textbf{\textit{(c)}} Reflectivity contrast for $GST-225$ at different laser fluences.\cite{sun}\textbf{\textit{(d)}} Crystal growth velocity $(v_g)$ for $GST-225$ at different laser fluences.\cite{behrens}\textbf{\textit{(e)}} XRD spectra for $GST-225$ at different temperatures. \textbf{\textit{(f)}} Crystallization fraction for $GST-225$ at different temperatures.\cite{sun} }
\end{figure}

Laser power is another important factor influencing the kinetics of $GST$ films that has been extensively studied in literature.\cite{kalb}\cite{wang}\cite{chen}\cite{chang} The evolution of crystallinity in the samples as the laser power is increased is clearly visible in the XRD measurements shown in Figure 8(a) \& 8(e).\cite{sun}\cite{behrens} It can be observed that even though a partial crystallization of th films can be observed for laser fluences between $15mJ/cm^2$ and $36mJ/cm^2$ in the optical reflectivity contrast measurements, this cannot be observed in the XRD spectra.\cite{sun} With the increase in the laser fluences from $36mJ/cm^2$ to $98mJ/cm^2$ a gradual but proportional increase in the volume of the crystalline fraction can be seen.\cite{sun} It is also seen that femtosecond pulses $(500fs)$ show a threshold power for crystallization at $13mJ/cm^2$ compared to $36mJ/cm^2$ for nanosecond pulses $(20ns)$, indicating the possibility for ultra-fast low power switching.\cite{sun}\cite{liu} Figure 8(b) shows the grain size as a function of laser power for both $ns$ and $fs$ pulses. \cite{sun}

Annealing of $GST$ films and the crystal growth velocity $(v_g)$ has also been studied extensively at different temperatures and laser fluences using both $DFT$ and experimental studies.\cite{behrens}\cite{ronenberger} It is apparent that the crystalline phase dominates the fraction of the solid as the temperature is increased and maintained above $T_g \sim 150^oC$ as seen from the XRD measurements and the crystal fraction measurements at different annealing temperatures.\cite{behrens}\cite{ronenberger} We can observe that $v_g$ increases with the laser power, this is to be expected with an increase in the rate of crystallization.\cite{behrens} The cooling rate also plays an important role in the crystallization kinetics, by saturating $v_g$ at high laser powers and promoting amorphization at high enough values.\cite{behrens} $v_g$ also increases with an increase in temperature till about $527^oC$ when the crystalline nucleus becomes too unstable to propagate.\cite{behrens} The theoretical maximum value for $v_g$ predicted by $DFT$ studies is $\sim 1.5m/s$ at $\sim 427^oC$ which is in close agreement with the experimentally measured value of $\sim 1.7m/s$.\cite{behrens}\cite{ronenberger}

Studies have also looked at how doping iso-electronic species, such as  other chalcogens like $Se$ affects the switching speed in $GST-225$, observing an improved performance in $Ge-Sb-Se-Te$ $(GSST)$, which similar to $GST$ forms a metastable cubic phase $(c-GSST)$ and a stable hexagonal phase $(h-GSST)$.\cite{vinod}\cite{guo} The optimized addition of $Se$ to $GST-225$ can significantly reduce the optical loss in the near-IR to mid-IR range, where $GSST$ shows a broadband transparency from $\lambda \sim 1 \mu m$ to $1.85\mu m$ while offering a $\Delta n \sim 2$ without loss penalty $(\Delta k \approx 0)$, maximizing the figure of merit $(FOM)=\Delta n/\Delta k$.\cite{vinod} The reason for this optical contrast according to $DFT$ simulations of random addition of $Se$ atoms to the $GST$ structure is the widening of the band gap leading to a smaller loss in the IR regime, which has been experimentally confirmed.\cite{luong}\cite{vinod} $GSST$ also shows a smaller density of states near the band edge due to the narrowing of the conduction and valence band in the $E-k$ diagram, which reduces the free carrier absorption in the IR range.\cite{vinod} The electronic excitations due to laser stimulation are also suppressed in $GSST$ compared to $GST$, where the creation of non-equilibrium photo-generated charge carriers are crucial for bond rupture.\cite{kolobov} Photo-generated carriers occupy these weak states in the longer $Ge-Te$ bond making them more susceptible to thermal vibration induced dislocations.\cite{kolobov}

It is also seen that increasing the $Se$ content monotonically increases the crystallization temperature, making the $c-GSST$ phase more stable and the crystallization process slower $(t\sim 3\mu$ at $25\% $ $Se)$ compared to $GST$ $(t\sim 20ns)$.\cite{wang}\cite{vinod} This also makes the effect of the preconditioning pulse weaker, pushing the system towards nucleation driven kinetics which is inherently sluggish.\cite{wang}\cite{vinod}

\begin{figure}[t]
\centering
\begin{subfigure}[b]{0.45\textwidth}
\centering
\includegraphics[width=1.7in]{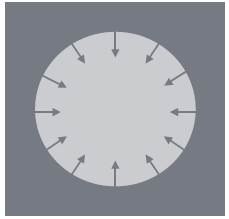}
\caption{{$AgIn$-doped $Sb_2Te$ and $Ge_{12}Sb_{88}$}}
\end{subfigure}
\begin{subfigure}[b]{0.45\textwidth}
\centering
\includegraphics[width=1.7in]{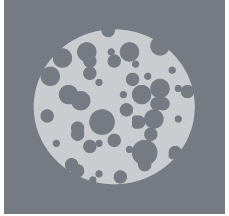}
\caption{{$Ge_1Sb_2Te_4$ and $Ge_2Sb_2Te_5$}}
\end{subfigure}

\caption{\textbf{\textit{(a)}} Rim to Centre Growth. \textbf{\textit{(b)}} Nucleation and subsequent Crystal Growth.\cite{kalb}}
\end{figure}

\subsection*{\normalfont Rim to Center Crystallization} 

One of the possible mechanisms for the $a-GST \rightarrow c-GST$ transformation to occur is rim to centre crystallization, where the crystal grows from the rim of the melted $a-GST$ spot toward the centre, hence its name.\cite{kalb} The primary reason for why this might occur would be if the edges of the spot cooled faster than the centre, then this would promote crystal growth in the inwards direction.\cite{kalb}\cite{welnic} This is only possible if heat is extracted fast enough and evenly at the rim-ambient interface and not the spot-substrate interface, forming a thermal gradient between rim and centre, like in the case of $AgIn-$doped $Sb_2Te_3$ where a direct correlation between the size of the spot and crystallization time is observed, characteristic of such a mechanism, confirmed by irradiation experiments of $AgIn-$doped $Sb_2Te_3$ observing the same between radius of the spot $(R_s)$ and crystallization time $(t_{r \rightarrow c}\sim 600ns)$.\cite{kalb} In such materials the crystal growth velocity $(v_g)$ is a exponential function of the temperature.\cite{kalb}\cite{behrens}\cite{ronenberger}  

\subsection*{\normalfont Crystal Nucleation \& Subsequent Growth}

The second, more probable mechanism for the $a-GST \rightarrow c-GST$ transformation occurring is nucleation followed by crystal growth.\cite{kalb}\cite{cai} Such a mechanism requires the laser irradiated $GST$ spot not losing heat fast enough to induce rim to centre growth, but instead causing the formation of randomly distributed crystal nucleation sites, facilitated by low thermal conductivity of the rim-ambient interface, spot-substrate interface and $GST$ alloys in general.\cite{wang} This is observed in $GST-415$, $GST-124$ and $GST-225$ where the crystallization time $t_n$ is independent of the radius of the spot, which is expected of such a mechanism.\cite{wang}\cite{liu}\cite{takeda} 

The rate at which nucleation sites are formed inside the irradiated spot is dependent on the temperature, which in-turn is determined by the laser power and pulse width.\cite{wang}\cite{behrens}\cite{ronenberger} This points at the existence of a optimum temperature for nucleation $(T \sim 175-180^oC)$ to occur, when crystallization is fastest in the time-temperature transformation $(t-T)$ diagram.\cite{wang}  The temperature dependence of the steady state homogeneous nucleation rate is strong and drops rapidly over a small temperature range. Thus when the laser power is sufficient to reach this optimum temperature, the crystallization is uniform and reproducible over a large number of cycles.\cite{wang}\cite{behrens}

Nucleation is generally accepted to be the time limiting step in such a process, making it the bottleneck for achieving fast switching speeds, especially when using as-deposited $a-GST$.\cite{wang} The crystallization of as-deposited $a-GST$ films is dominated by nucleation, making the $a-GST \rightarrow c-GST$ orders of magnitude slower compared to melt-quenched $GST$ $(t\sim 3\mu m)$, which contain sub-critical crystalline domains left over from when $c-GST$ undergoes the laser amorphization process. These domains allow us to bypass nucleation and proceed with the crystal growth dominated step, making the process significantly faster $(t\sim 20 ns)$.\cite{wang}\cite{zhang}\cite{cai} Higher laser powers than the optimum often lead to a random distribution of nucleation centres, but also a higher probability of ablation which is undesirable and results in the loss of information stored in the memory cell.\cite{zhang}\cite{qi}\cite{zhou}\cite{han}

Crystallization time may also be much larger than the duration of the laser pulse in cases where the substrate on which the $GST$ spots are made has a cooling time much larger than the pulse itself, leading to the slow cooling and better crystallization of the $GST$.\cite{kalb}\cite{chenbin}\cite{welnic} This usually happens when the thermal conductivity of the substrate is significantly less compared to the $GST$, for example $GST$ films on $SiO_2$ and $Al_2O_3$ substrates which show a high degree of crystallinity due to the cooling time for $SiO_2$ and $Al_2O_3$ being of the order of nanoseconds $(t\sim 10ns)$.\cite{kalb}\cite{welnic} Thus the cooling time for the substrate is an important factor which limits the laser pulse widths that can be utilized for crystallization.\cite{kalb}\cite{khulbe}\cite{welnic}\cite{wang}

\section*{\normalfont Atomistic Mechanism of Phase Transformation}

Now that we have looked into the various phases of $GST-225$ as well as the macroscopic kinetics of the phase transformation upon optical stimulation, we turn our attention to the atomic mechanism for the $c-GST \rightarrow a-GST$ transformation. 

\subsection*{\normalfont Atomic Migrations \& Bond breaking}
It is evident from our earlier description of the $c-GST$ unit cell, which crystallizes in a distorted rock-salt cubic structure, that there exist three long $(\sim 3.20$\AA $)$  and three short $(\sim 2.84$\AA $)$ $Ge-Te$ bonds.\cite{tsafack}\cite{kolobov} This in-turn results in weaker $(E_b\sim 396.7 kJ/mol)$  and stronger $(E_b\sim 402kJ/mol )$ $Ge-Te$ bonds, determined by their bond dissociation energies.\cite{coombs}\cite{kolobov}\cite{cotrell} Upon laser stimulation, the longer bonds are broken but not the shorter bonds as a result of which the $Ge$ in the distorted octahedral centre is pulled into a tetrahedral void in one of the corners by the shorter $Ge-Te$ bonds as shown in Figure 6(b).\cite{kolobov}\cite{zhitang} This is clear as the measured co-ordination number for $Ge$ in $a-GST$ $\approx 4$ and the simulated XANES spectra for this process and the measured data are in good agreement as seen in Figure 7(b) \& (c).\cite{kolobov} As the direction in which these longer and shorter bonds exist depends on the surrounding atomic layers , and the $8$ fold $3D$ symmetry of a cubic unit cell adds even more randomness as to which of the four tetrahedral centres the $Ge$ will be pulled towards, this creates a loss of long range order in the crystal and causes amorphization.\cite{wang}\cite{kolobov} The train of thought that has been presented herein is also applicable to a octahedral centre containing a $Sb$ atom where the longer $(\sim 3.10$\AA$)$ and shorter $(\sim 2.95$\AA$)$ $Sb-Te$ bond display a similar behaviour.\cite{kolobov}\cite{zhitang}\cite{sun}

\subsection*{\normalfont Peierl's Distortion \& Umbrella Flip Distortion}

Though the $c-GST \rightarrow a-GST$ transformation involves the longer $Ge-Te$ bond rupturing upon laser stimulation, it is important to note that the solid can not be considered conventionally molten due to the shorter $Ge-Te$ bonds remaining unbroken.\cite{wang}\cite{kolobov}\cite{zhitang} Thus the overall transformation from $c-GST \rightarrow a-GST$ can be thought of as a flip of the $Ge/Sb$ atom from a octahedral position to a tetrahedral position, which involves the rupture of the longer bonds and a direction change of the shorter bonds similar to the ribs of an  umbrella flipping during strong winds.\cite{tsafack}\cite{kolobov} This type of distortion is therefore aptly named the umbrella flip distortion.\cite{kolobov} In actuality it is more similar to a localized Peierls distortion of the lattice, which involves atoms in a chain coming close to one of their neighbours, which in this case would be the $Ge$ atom coming close to the $Te$ atoms on the face-centre and the corner as seen in Figure 6(b).\cite{kolobov}\cite{zhitang} These distortions of the lattice are the primary reason for the change in the band gap of the solid upon stimulation, the loss in the long range order and formation of dangling bonds also results in the formation of an Urbach-Martienssen tail in their absorption spectra characteristic of amorphous materials as seen in Figure 3(d).\cite{shportko}\cite{kellner}

\begin{figure}[]
\centering
\begin{subfigure}[b]{1\textwidth}
\centering
\includegraphics[width=4.9in]{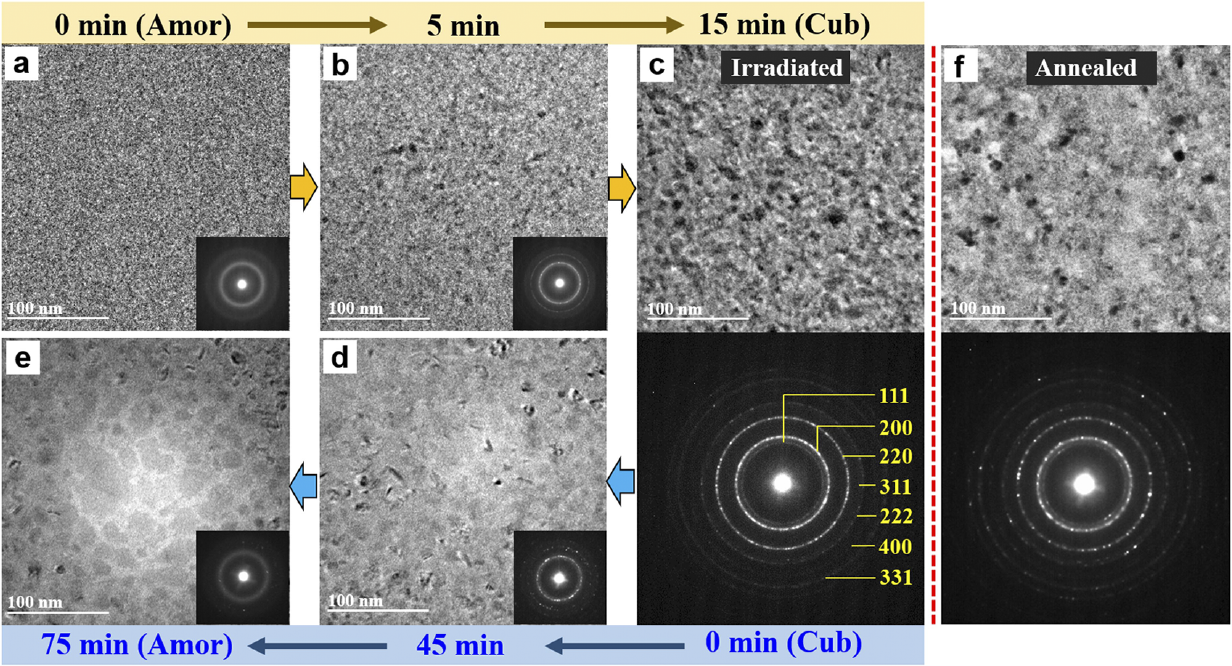}
\caption{{}}
\end{subfigure}
\begin{subfigure}[b]{1\textwidth}
\centering
\includegraphics[width=4.9in]{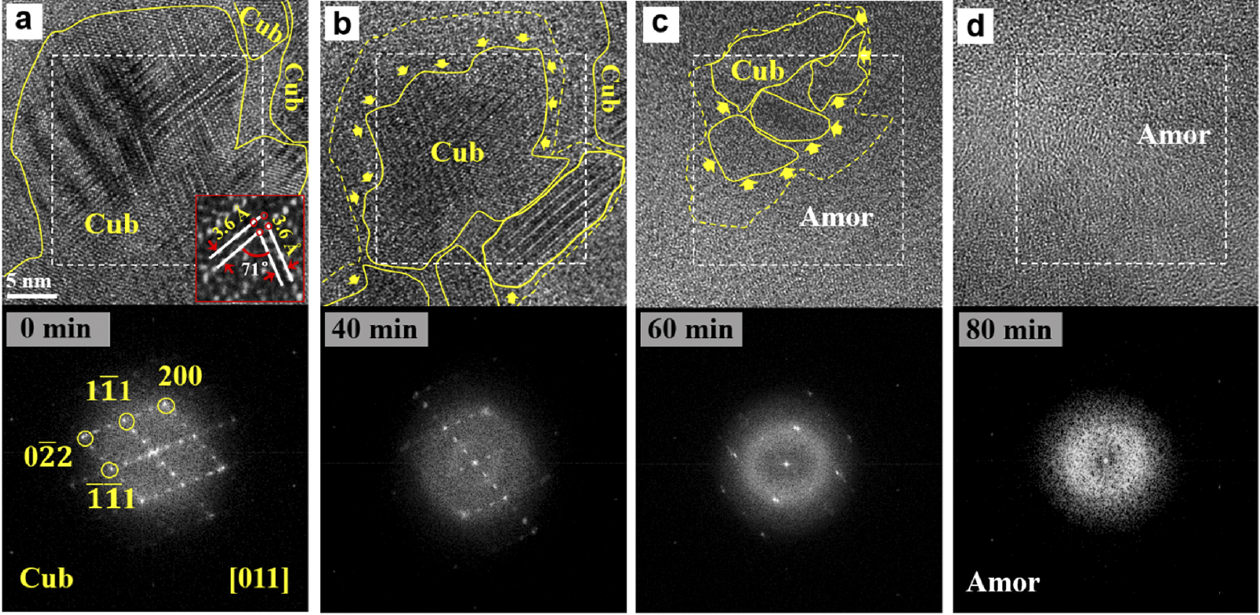}
\caption{{}}
\end{subfigure}

\caption{\textbf{\textit{(a)}} $c \rightarrow a$ and $a \rightarrow c$ process under a beam intensity of $6$ x $10^{23} e^-/m^2s$. \textbf{\textit{(b)}} $c \rightarrow a$ process under a beam energy of $200keV$.\cite{jiang}}
\end{figure}

\subsection*{\normalfont Effect of charge carriers}

The laser pulse not only heats the material but also excites non equilibrium charge carriers that weaken the longer $Ge-Te$ bonds significantly.\cite{wang}\cite{jiang} Since the density of states in these bonds is lower, the carriers more easily populate these states making hem more vulnerable to thermal vibrations and dislocations.\cite{tsafack}\cite{kellner}\cite{vinod} Thus the presence of these electronic interactions is critical to the phase transformation. This can be most clearly demonstrated by a study where a similar phase change was observed when a electron beam was used instead of a laser, indicating that carrier interactions with the lattice play a major role in the overall process.\cite{jiang}
The study conducted employed a $200 keV$ electron source with a electron density of  $6$ x $10^{23} e^-/m^2s$ which was used to irradiate a $80nm$ $a-GST$ film.\cite{jiang} The films were exposed to the beam for $5$ $min$ when multiple crystal domains began to appear which were confirmed by SAED. On increasing he exposure time to $15$ $min$ th crystallite size grows to $10-20nm$ and SAED shows sharp and bright diffraction rings.\cite{jiang} The study also confirmed the existence of a threshold beam intensity $(1.6$ x $10^{23}e^-.m^2s)$ below which no transformation was observed however long the beam remains irradiated on the sample.\cite{jiang} Amorphization is also possible by increasing the beam intensity to about $1.1 $x $10^{24}e^-/m^2s$.\cite{jiang} This was confirmed as being due to the electron beam by performing the same experiments on thermally amorphized samples.\cite{jiang} The threshold intensity for amorphization was found to be $8$ x $10^{23}e^-.m^2s$. This radiation amorphized $GST$ can be recrystallized by adjusting the electron beam intensity.\cite{jiang} Irradiation induced amorphization was also found to be continuous with no abrupt changes in the structural patterns, which indicates a non thermal component to th phase change. \cite{jiang} The temperature increase due to irradiation is calculated using the expression $\Delta T=\frac{{1}}{\pi \kappa e}\frac{\Delta E}{d}\ln\frac{b}{a}$, which gives us a result around $220^oC$ nowhere close to $\sim 650^oC$ needed for thermal amorphization.\cite{jiang} It was also observed that the intensity of the electron beam, voltage, and irradiation time had a major impact on the degree of phase change that was observed in the sample, making us conclude that the collisions of the electrons with the lattice plays an important role in triggering $a \rightarrow c$ transformation, which can be localized by decreasing beam area, additionally electronic states also change from $p-type$ bonding to a $sp^3$ thus leading to optical contrast between the two phases.\cite{kolobov}\cite{kellner}\cite{jiang} In order for the absorbed energy to be dissipated non-radiatively, a large number of phonon's have to be emitted which is facilitated by the presence of  large number of atomic sites near these states, enabling us to selectively heat these areas to a higher temperature.\cite{siegert}\cite{wang}\cite{mukhopadhyay}


\section*{\normalfont Conclusion}

In this term paper, we have investigated the $c \rightarrow a$ ans $a \rightarrow c$ transformation of $GST-225$ under optical stimulation by a laser source. We first looked at the three structural phases of $GST$ namely $c-GST$ (rock-salt), $h-GST$ (hexagonal) and $a-GST$ (amorphous),  their band structures, vacancy distributions and atomic and bond lengths and arrangements which play a major role in their stability, phase transformation as well as electrical and optical contrast. Then the role of laser power, pulse duration, temperature, doping as well as the distribution of electronic states and carriers and their effect on the macroscopic kinetics and the mechanism of phase transformation was explored. It was found out that the transformation prefers a nucleation mechanism followed by subsequent crystal growth due to the combined thermal characteristic of both the $GST$ and the substrate, as well as laser power and pulse width.

Then we delved into the atomic mechanism of the phase transformation using both experimental evidence from literature in tandem with $DFT$ calculations to figure out the cause for the transformation. This analysis points to the primary cause for the $c \rightarrow a$ transformation being the atomic migration of the $Ge/Sb$ atoms from the octahedral to the tetrahedral centre, causing a loss in the long range order in the arrangement i.e. amorphization. The rupture of the longer $Ge-Te/ Sb-Te$ bonds, which leads to a umbrella-flip distortion (Peierls distortion) pulling the $Ge/Sb$ from a $6$- fold co-ordination $(octahedral)$ site to $4$-fold co-ordination $(tetrahedral)$ site in a semi-random manner, was found to be the root cause. The importance of the role played by charge carriers and their interactions with the lattice in causing the phase transformation is also highlighted.
In conclusion, we have attempted to provide a comprehensive picture of the macroscopic kinetics of the phase switching behaviour observed in $GST-225$ and the parameters that affect the stability of the phases as well as the speed of the phase switching. We also present a atomistic picture of the phase transformation by studying the interactions between the constituent atoms, their movement and behaviour in the crystal structure to pinpoint as reason for the phase change behaviour.
\bibliography{myref}
\bibliographystyle{unsrt}

\end{document}